\documentclass[reprint,prd,superscriptaddress,showpacs,nofootinbib,floatfix,preprintnumbers]{revtex4-1}
\usepackage{graphicx, bm, color, xcolor, amsmath}

\begin{document}
\preprint{MAN/HEP/2014/14}
\title{
A Possible Two-component Flux for the High Energy Neutrino Events at IceCube }
\author{Chien-Yi Chen}
\affiliation{Department of Physics, Brookhaven National Laboratory, Upton, New York 11973, USA}

\author{P. S. Bhupal Dev}
\affiliation{Consortium for Fundamental Physics, School of Physics and Astronomy, University of Manchester, Manchester M13 9PL, United Kingdom}

\author{Amarjit Soni}
\affiliation{Department of Physics, Brookhaven National Laboratory, Upton, New York 11973, USA}
\date{\today} 
\begin{abstract} 
Understanding the spectral and flavor composition of the astrophysical neutrino flux responsible for the recently observed ultra-high energy events at IceCube is of great importance for both astrophysics and particle physics. We perform a statistical likelihood analysis to the 3-year IceCube data and derive the allowed range of the spectral index and flux normalization for various well-motivated physical flavor compositions at source. While most of the existing analyses so far assume the flavor composition of the neutrinos at an astrophysical source to be (1:2:0), it seems rather unnatural to assume only one type of source, once we recognize the possibility of at least two  physical sources. Bearing this in mind, we entertain the possibility of a two-component source for the analysis of IceCube data. It appears that our two component hypothesis explains some key features of the data better than a single-component scenario, {\it i.e} it addresses the apparent energy gap between 400 TeV to about 1 PeV and easily accommodates the observed track to shower ratio. 
Given the extreme importance of the flavor composition for the correct interpretation of the underlying astrophysical processes as well as for the ramification for particle physics, this two-component flux should be tested as more data is accumulated.  
\end{abstract}
\maketitle
%%%%%%%%%%%%%%%%%%%%%%%%%%
\section{Introduction} \label{sec:1}
%%%%%%%%%%%%%%%%%%%%%%%%%%%
%{\em Introduction.---}
The recent observation of ultra-high energy (UHE) neutrino events at IceCube~\cite{ic1, ic2, ic3} in previously uncharted energy regime has commenced a new era in Neutrino Astrophysics. 
Following the initial two events around 1 PeV deposited energy~\cite{ic1}, additional 26 events were found in the 30 - 400 TeV energy range~\cite{ic2} with the 2-year dataset. More recently, further 9 events were reported with the 3-year dataset~\cite{ic3}, with one event at 2 PeV, the highest-energy neutrino interaction ever observed in Nature. Together, the observed total of 37 candidate events reject a purely atmospheric explanation at $5.7\sigma$~\cite{ic3} and strongly suggest an extra-terrestrial origin. This provides a unique opportunity to {\em directly} probe the energetic physical processes occurring in dense astrophysical environments, which are otherwise inaccessible with traditional messengers like photons or charged particles in cosmic rays. 

It is imperative for both astrophysics and particle physics to understand all possible aspects of the UHE neutrino events, and in particular, to extract information on the possible source(s) and the underlying spectral shape of the astrophysical neutrino flux (for reviews, see e.g.~\cite{Anchordoqui:2013dnh, Murase:2014tsa}). 
Since no significant clustering is observed~\cite{ic3} and there is no evidence for point-like sources of astrophysical neutrinos~\cite{ic5}, the  current data suggests either many isotropically distributed point sources or some spatially extended sources. Moreover, most of the UHE neutrino events have arrival directions in high galactic latitudes~\cite{ic3}, thereby suggesting a dominant extragalactic component~\cite{Anchordoqui:2013qsi}, %, Wang:2014jca}, 
which could be attributed to various astrophysical sources.\footnote{A sub-dominant galactic contribution, possibly associated with known local large diffuse TeV to PeV $\gamma$-ray sources at the galactic center~\cite{Fox:2013oza} %, Neronov:2013lza, Razzaque:2013uoa, Ahlers:2013xia, Supanitsky:2013ooa, Bai:2014kba, Anchordoqui:2014rca} (including the Fermi bubbles)%~\cite{Ahlers:2013xia, Razzaque:2013uoa, Lunardini:2013gva, Taylor:2014hya} 
or the interstellar medium~\cite{Taylor:2014hya} cannot be ruled out yet. } Typical examples are cosmic-ray (CR) reservoirs like star-burst galaxies and galaxy clusters/groups~\cite{Murase:2013rfa}, %, Liu:2013wia, Chang:2014hua, Anchordoqui:2014yva, Tamborra:2014xia, Zandanel:2014pva},  
CR accelerators like active galactic nuclei (AGNs)~\cite{Essey:2009ju, Murase:2014foa}, %Padovani:2014bha, Krauss:2014tna, Tavecchio:2014iza, Sahu:2014fua, Dermer:2014ata, Kalashev:2014vya, Dermer:2014vaa} 
% (including blazars and quasars),  
gamma-ray bursts~\cite{Petropoulou:2014lja} %, Dado:2014mea, Fargion:2014jaa, Bustamante:2014oka} 
and newborn pulsars~\cite{Murase:2009pg}, or even charmed meson decays in mildly relativistic jets of supernovae~\cite{Bhattacharya:2014sta}. A cosmogenic source due to UHECR interactions with the CMB background~\cite{GZK} is now disfavored~\cite{Sigl:2014jna}. 

%In the IceCube analyses so far, the astrophysical neutrinos are assumed to have originated from the decay of charged pions/kaons produced through hadro-nuclear ($pp$) or photo-hadronic ($p\gamma$) interactions of UHECRs in a dense astrophysical system. 
There are two conventional production sources of UHE neutrinos from interactions of UHECRs in a dense astrophysical system~\cite{gaisser}, namely, (i) hadro-nuclear production by inelastic $pp$ or $pn$ scattering in cosmic-ray reservoirs like starburst galaxies and galaxy clusters/groups, and (ii) photo-hadronic production by $p\gamma$ scattering in cosmic ray accelerators like GRBs and AGN. Both kinds of sources produce charged pions/kaons, whose subsequent decays are expected to give rise to astrophysical neutrinos. For charged pions produced by $pp$ scattering, isospin invariance yields a roughly equal ratio of $\pi^+,~\pi^-$ and $\pi^0$ production, and the subsequent decay chain 
\begin{align}
\pi^\pm \to \mu^\pm+\nu_\mu(\bar{\nu}_\mu), \quad \mu^\pm \to e^\pm + \nu_e(\bar{\nu}_e)+\bar{\nu}_\mu (\nu_\mu)
\label{pidecay}
\end{align} 
leads to a flavor composition of ($\nu_e$:$\nu_\mu$:$\nu_\tau$)$_{\rm S}$=(1:2:0)$_{\rm S}$ at the source (S). Note that the kinematics of the decay chain  is such that each neutrino in the decay chain carries off roughly equal energy~\cite{gaisser}. After 
the neutrino oscillations are averaged over an astronomical distance scale, the final composition on Earth (E) becomes (1:1:1)$_{\rm E}$~\cite{Learned:1994wg} for a tri-bi-maximal (TBM) neutrino mixing pattern~\cite{Harrison:2002er}. The flux of neutrinos coming from $pp$ collisions follows that of the progenitor protons, which is typically a power-law spectrum $\Phi \propto E^{-\gamma}$ with $\gamma\sim 2$ for a diffusive Fermi-shock acceleration mechanism, whereas the neutrino flux due to $p\gamma$ collisions has a strong $\Delta^+$ resonance peak, and therefore, falls off at lower energies~\cite{gaisser}.  

Using the (1:1:1)$_{\rm E}$ flavor composition and assuming a single-component $E^{-\gamma}$ flux over the entire energy range of interest, it was shown~\cite{Chen:2013dza, Laha:2013lka} that the 2-year IceCube data was largely consistent with the expectations from the Standard Model (SM) neutrino-nucleon interactions. % within the known uncertainties. 
This provides a unique test of the SM involving the highest energy neutrinos ever observed in Nature and any statistically significant deviations in future might call for a non-standard explanation. In fact, several New Physics scenarios have been envisaged in this context, e.g.~early-decay of a massive long-lived particle~\cite{Ema:2013nda}, decay~\cite{Feldstein:2013kka}  or annihilation~\cite{Zavala:2014dla} of a heavy Dark Matter, secret neutrino interactions involving a light mediator~\cite{nuSI}, lepto-quark resonance~\cite{Barger:2013pla}, decay of massive neutrinos to light ones over cosmological distances~\cite{Baerwald:2012kc}, mirror neutrinos~\cite{Joshipura:2013yba}, superluminal 
neutrinos~\cite{Borriello:2013ala}, color-octet neutrinos~\cite{Akay:2014tga}, extra-dimensions~\cite{Aeikens:2014yga} and TeV-scale gravity~\cite{Illana:2014bda}.  
%Most of these proposals are motivated by some specific features in the IceCube 
%dataset such as an apparent paucity of muon tracks, a possible energy gap between 300 TeV and 1 PeV, and the absence of events above PeV energies. 
%For a reliable interpretation of the IceCube data, 
%it is however  Note that 
Even within the SM framework, various other possibilities have been considered, e.g. the Glashow resonance in $\bar\nu_e e^-$~\cite{Bhattacharya:2012fh} and $\nu_\ell\gamma$ scattering~\cite{Alikhanov:2014uja}, %anomalous radiative transitions involving photons~\cite{Ishikawa:2014uma} 
and interactions of nuclei with matter~\cite{Winter:2014pya}. If the data continues to be consistent with  the SM predictions, one can put useful constraints on some of the exotic scenarios mentioned above~\cite{Murase:2014tsa}, which are otherwise difficult to probe in low-energy laboratory experiments. 

Nevertheless, it was pointed out in~\cite{Chen:2013dza} that the (1:1:1)$_{\rm E}$ flux seems to give a mild deficit in the observed muon tracks at high energies, where the atmospheric background is anyway expected to be small. This was independently confirmed  in a dedicated likelihood analysis~\cite{Mena:2014sja}. Although the (1:1:1)$_{\rm E}$ flavor composition is still allowed at the 90\% C.L.~\cite{Mena:2015mka, Aartsen:2015ivb} and a potential muon deficit could be attributed to experimental effects such as track mis-identification, it is worth scrutinizing other possible flavor ratios at source. 
%, which mildly disfavors the (1:1:1)$_{\rm E}$ composition. %at 92\% CL using the 3-year data. Ref.~\cite{Mena:2014sja} argued that (1:0:0)$_{\rm E}$ provides the best-fit, but this composition cannot be attained from {\em any} flavor ratio at an astrophysical source within the standard neutrino oscillation framework~\cite{pdg}. If confirmed, this would suggest either a misunderstanding of the expected background or a misidentification of tracks or some New Physics. 
%In light of this, 

In this paper, we critically examine the possible physical flavor and spectral compositions of the UHE neutrino flux in light of the 3-year IceCube data. After confirming the mild `muon deficit problem' with the standard (1:2:0)$_{\rm S}$ flux, we consider other well-motivated sources with (1:0:0)${\rm _S}$,  (0:1:0)$_{\rm S}$ and (1:1:0)$_{\rm S}$ flavor compositions and show that it is possible to mitigate the muon deficit problem, if it really exists, with the (1:0:0)${\rm _S}$ flux, whereas the (0:1:0)$_{\rm S}$ and (1:1:0)$_{\rm S}$ fluxes further aggravate the problem. In any case, once one recognizes the existence of any of these additional sources along with the standard (1:2:0)$_{\rm S}$ source, it is rather {\em natural} to consider at least a {\em two-component} flux, instead of a single component over the entire energy range of interest. Moreover, we show that such a two-component flux could also offer a simple explanation of the apparent energy gap between 400 TeV--1 PeV, apart from addressing the muon deficit problem, all {\em within} the  SM framework, i.e.~without invoking any New Physics. 

The plan of the paper is as follows: in Section~\ref{sec:2}, we give the pertinent details of the calculation of event rate at IceCube, as used in our simulation. In Section~\ref{sec:flav}, we discuss various possible flavor compositions at source. In Section~\ref{sec:3}, we perform a Poisson likelihood analysis for different astrophysical flavor compositions and flux normalizations. %In Section~\ref{sec:4}, we discuss the possibility of an astrophysical source with (1:0:0)$_{\rm S}$ flavor composition. 
In Section~\ref{sec:5}, we analyze the two-component flux with various flavor compositions. Our conclusions are given in Section~\ref{sec:6}. 

%%%%%%%%%%%%%%%%%%%%%%%%%%%%%%%%%%%%%%%%%%%%%%%%%
\section{Event Rate} \label{sec:2}
%%%%%%%%%%%%%%%%%%%%%%%%%%%%%%%%%%%%%%%%%%%%%%%%%

%{\em Event Rate.---}
The expected number of neutrino-induced events at IceCube %as a function of the deposited energy 
can be written as 
\begin{align}
N \  = \ TN_A\Omega \int_{E_{\rm min}}^{E_{\rm max}}dE_{\rm dep} \int_0^1 dy~ \Phi \: V_{\rm eff}\: A\: %\nonumber \\
%& \qquad \qquad \qquad \times  
\frac{d\sigma}{dy} ,
\label{eq:N}
\end{align}
where $E_{\rm dep}$ is the electromagnetic-equivalent deposited energy, which is {\it always} smaller than the incoming neutrino energy $E_\nu$ in the laboratory frame by a factor depending on $E_\nu$ and the type of interaction, $T$ is the time of exposure, $N_A$ is the Avogadro number, $\Omega$ is the solid angle of coverage, $\Phi$ is the incident neutrino flux, $V_{\rm eff}$ is the effective target volume of the detector, $A$ is the attenuation factor for upgoing neutrinos traveling through the Earth material, $\sigma$ is the neutrino-induced interaction cross section, and $y=(E_\nu-E_\ell)/E_\nu$, $E_\ell$ is the inelasticity parameter which is a measure of the energy carried by the outgoing lepton in the laboratory frame.  
%is the inelasticity parameter which is a measure of the energy transferred between lepton and hadron systems in the neutrino-nucleon interaction. 
%Note that the differential cross section in (\ref{eq:N}) is written only in terms of $y$, and has been integrated over the Bjorken scaling variable $x$ for neutrino-nucleon interactions. 
%where $-Q^2$ is the invariant momentum-square transfer to the mediator and $M_N$ is the mass of the 
%since none of the other parameters appearing in (\ref{eq:N}) do not depend on $x$. 
The limits of the energy integration $E_{\rm min}$ and $E_{\rm max}$ give the bin size over which the expected number of events is being calculated. 

The numerical values of the various parameters appearing in Eq.~(\ref{eq:N}) as used in our analysis are computed using the procedure given below: 
\begin{enumerate}
\item [(i)] $T=988$ days for the IceCube data collected between 2010--2013~\cite{ic3}. 
\item [(ii)]
 $N_A = 6.022\times 10^{23}\:{\rm mol}^{-1}$, which is equal to $6.022\times 10^{23}\:{\rm cm}^{-3}$ water equivalent for interactions with the ice nuclei. For interactions with electrons, $N_A$ should be  replaced with $(10/18)N_A$ for the number of 
electrons in one mole of H$_2$O. 
\item [(iii)] $\Omega=4\pi\:{\rm sr}$ for an isotropic 
neutrino flux. For the upgoing events,~i.e.~those coming from the northern hemisphere at IceCube, 
we must include an attenuation factor due to scattering within the Earth, which is  represented by an energy-dependent shadow factor
~\cite{gandhi}
\begin{eqnarray}
A(E_\nu) \ = \ \frac{1}{2}\int_{-1}^1 d(\cos\theta)~{\rm exp}\left[-\frac{z(\theta)}{L_{\rm int}(E_\nu)}\right], 
\end{eqnarray}
where $\theta$ is the incident angle of the incoming neutrinos above nadir, $L_{\rm int}(E_\nu)=1/\sigma(E_\nu) N_A$ is the interaction length, and $z(\theta)$ is the effective column depth for upgoing neutrinos (for downgoing events, $z(\theta)=0$), which is obtained from the Earth density profile as given by the 
Preliminary Reference  Earth Model~\cite{prem}. The Earth attenuation effects become important at energies above $\sim 100$ TeV, making the Earth opaque to UHE neutrinos~\cite{gandhi}. This is why all the PeV events observed so far at IceCube are downgoing events. 
For the upgoing $\tau$-neutrinos, one should also include the regeneration effects inside the Earth~\cite{Dutta:2000jv}, which lead to fast $\tau$-decays producing 
secondary neutrinos of all flavors with lesser energy than the original incident one~\cite{Beacom:2001xn}.\footnote{Our estimates show that the regeneration effect on the total number of events in the energy range of interest is $\lesssim 5$\%.} 
\item [(iv)] $V_{\rm eff}(E_\nu)=M_{\rm eff}(E_\nu)/\rho_{\rm ice}$  is the effective target volume, where $\rho_{\rm ice}=0.9167\:{\rm g\:cm}^{-3}$ is the density of natural ice and $M_{\rm eff}$  is the effective target mass which includes the background rejection cuts and event containment criteria~\cite{ic2}. 
$M_{\rm eff}$ depends on the incoming neutrino energy and attains its maximum value $M_{\rm eff}^{\rm max}\simeq 400\:{\rm Mton}$, corresponding to $V_{\rm eff}^{\rm max}\simeq 0.44\:{\rm km}^3$ water-equivalent, above 100 TeV for $\nu_e$ CC events, and above 1 PeV for other CC and NC events~\cite{ic2}. There is some flavor bias at low energies caused by the deposited energy threshold due to missing energy in escaping particles from $\nu_\mu$ and $\nu_\tau$ CC events as well as all flavor NC events, which decreases $M_{\rm eff}$ for these events as compared to the $\nu_e$ CC events. 
\item [(v)] For the incoming neutrino flux, we first assume a single-component unbroken power-law spectrum: 
\begin{eqnarray}
\Phi (E_\nu) \ =\  \Phi_0 \left(\frac{E_\nu}{E_0}\right)^{-\gamma} ,
\label{phi}
\end{eqnarray}
where $\Phi_0$ is the {\em total} $\nu+\bar{\nu}$ flux for all flavors at $E_0=100$ TeV in units of ${\rm GeV}^{-1}{\rm cm}^{-2}{\rm sr}^{-1}{\rm s}^{-1}$ and $\gamma$ is the spectral index.\footnote{In the notation followed by our earlier analysis~\cite{Chen:2013dza}, $\Phi = CE^{-\gamma}$, where $C=(10^{10}~{\rm GeV}^2)\Phi_0 E_0^{\gamma-2}$.} 
The exact energy dependence might vary for different extra-terrestrial source evolution models~\cite{Murase:2014tsa}, and hence, the spectral index $\gamma$ is kept as a free parameter in our analysis. 
\item [(vi)] $E_{\rm dep}$ as a function of $E_\nu$  is calculated using the procedure outlined in~\cite{Chen:2013dza}; see also~\cite{Aartsen:2013vja, Kistler:2013my, Mena:2015mka}. For the cascade events caused by $\nu_e,~\nu_\tau$ charged-current (CC) and a sub-dominant all-flavor neutral current (NC) interactions, the underlying true neutrino energy can be reconstructed better than that for the track events caused by $\nu_\mu$ CC interactions~\cite{candia}. In the latter case, the true muon neutrino energy could be much higher than the deposited energy due to the through-going muons, thus allowing us to set only a {\em lower} limit on $E_\nu$~\cite{Aartsen:2013vja}. 
%The deposited energy $E_{\rm dep}$ in Eq.~(\ref{eq:N}) depends on the incoming neutrino energy $E_\nu$ as well as the type of interaction. For neutral current interactions $\nu_\ell+N \to \nu_\ell+X$, where $X$ is the outgoing hadron carrying energy $yE_\nu$, the deposited energy is given by $E_{\rm dep, had}= F_XyE_\nu$, where $F_X$ is the conversion efficieny of the photo-electrons for a hadronic shower~\cite{Kowalski:2004qc, Gabriel:1993ai}. For charged current interactions $\nu_\ell+N \to \ell+X$, the hadronic showers deposit an energy $E_{\rm dep, had}$ as given above, whereas for $\ell=e$, the electromagnetic shower deposits an energy $E_{{\rm dep},e}=(1-y)E_\nu$.  For $\ell=\mu$, we can only 
\item [(vii)] The (anti)neutrino-nucleon cross sections are calculated using the {\tt NNPDF2.3}~\cite{Ball:2012cx} parton distribution functions (PDFs) at next-to-next-to-leading order. With $x$-grids as low as $10^{-9}$ and $Q^2$ grids up to $10^8~{\rm GeV}^2$, they have a relatively small error on the cross sections for the current energy range of interest~\cite{Chen:2013dza}. 
For a more precise determination of the cross sections at high energies, one has to include the non-linear QCD effects~\cite{Brambilla:2014jmp} beyond the DGLAP formalism~\cite{dglap}. 
\end{enumerate}
%%%%%%%%%%%%%%%%%%%%%%%%%%%%%%
\section{Flavor Composition} \label{sec:flav}
Given a flavor ratio ($f_e^0$:$f_\mu^0$:$f_\tau^0$)$_{\rm S}$ of $\nu+\bar{\nu}$ at source, the corresponding value ($f_e$:$f_\mu$:$f_\tau$)$_{\rm E}$ on Earth is given by
\begin{eqnarray}
f_\ell \ = \ \sum_{\ell'=e,\mu,\tau} \sum_{i=1}^3 |U_{\ell i}|^2|U_{\ell' i}|^2 f_{\ell'}^0  \ \equiv \ \sum_{\ell'} P_{\ell \ell'} f_{\ell'}^0 \; ,
\label{flavor}
\end{eqnarray} 
where $U_{\ell i}$ are the elements of the PMNS mixing matrix and $P_{\ell \ell'}$ is the oscillation probability for $\nu_\ell \to \nu_{\ell'}$ in vacuum. Assuming a TBM mixing~\cite{Harrison:2002er}, which is a good approximation at this stage~\cite{Fu:2012zr}, the positive definite elements of $P$ are given by 
\begin{align}
P \ = \ \frac{1}{18}\left(\begin{array}{ccc} 
10 & 4 & 4 \\
4 & 7 & 7 \\
4 & 7 & 7
\end{array}
\right) \, . 
\label{tbm}
\end{align}
Using this in Eq.~\eqref{flavor}, one can easily check that a (1:2:0)$_{\rm S}$ flavor composition at the source [cf.~Eq.~\eqref{pidecay}] will lead to a (1:1:1)$_{\rm E}$ composition on Earth.\footnote{Using the current $3\sigma$ values of the neutrino mixing parameters~\cite{pdg} instead of the TBM structure \eqref{tbm}, we obtain (0.9-1:1.1-1:1-0.9)$_{\rm E}$ which can be safely assumed to be (1:1:1)$_{\rm E}$ for numerical purposes.}  Here we make an interesting observation that for an observed (1:1:1)$_{\rm E}$ composition on Earth, the source composition may not be necessarily (1:2:0)$_{\rm E}$, as a flavor-universal source composition of (1:1:1)$_{\rm S}$ also leads to the same composition  (1:1:1)$_{\rm E}$ on Earth.

Apart from the standard $\pi^\pm$ production mode \eqref{pidecay} giving rise to (1:2:0)$_{\rm S}$ flavor composition of astrophysical neutrinos, there are a few other possibilities~\cite{flav1}, leading to different flavor compositions as follows: 
\begin{itemize}
\item [(i)] If the charged pions are produced by $p\gamma$ scattering, the $\Delta^+$ resonance gives rise to $\pi^++n$ and $\pi^0+p$, in the ratio of 1:2. Since $\pi^-$ (and hence $\mu^-$ and $\bar{\nu}_e$) production is suppressed in this case, we get the source flavor ratio (1:1:0)$_{\rm S}$ for neutrinos and  (0:1:0)$_{\rm S}$ for anti-neutrinos. Using Eqs.~\eqref{flavor} and \eqref{tbm}, we get the corresponding earthly flavor ratios of (14:11:11)$_{\rm E}$ and (4:7:7)$_{\rm E}$ respectively. Note that (1:1:0)$_{\rm S}\equiv $ (14:11:11)$_{\rm E}$ flavor ratio for both neutrinos and antineutrinos is also possible in prompt decays of charmed particles. 
\item [(ii)] Since the rest-frame lifetime of muons, $\tau_\mu=2.2\times 10^{-6}$ s, is larger than that of charged pions, $\tau_{\pi^\pm}=2.6\times 10^{-8}$ s~\cite{pdg}, it is possible that the muon in the decay chain of Eq.~\eqref{pidecay} loses energy in the source environment, e.g. due to synchrotron radiation in a strong magnetic field or by scattering in a dense astrophycal medium, before decaying~\cite{Rachen:1998fd, Kashti:2005qa}. In this case, the flavor composition at source will be (0:1:0)$_{\rm S}$ and the corresponding earthly composition will be (4:7:7)$_{\rm E}$. 

\item [(iii)] It is also possible to have a purely electron (anti)neutrino flux at source, i.e. with flavor ratio (1:0:0)$_{\rm S}$, which will give rise to (5:2:2)$_{\rm E}$ flux on Earth. This is possible e.g. when the source injects a nearly pure neutron flux~\cite{Anchordoqui:2003vc, Anchordoqui:2014pca}. 
%This will be discussed in more details in Section~\ref{sec:4}.  
\end{itemize}

%%%%%%%%%%%%%%%%%%%%%%%%%%%%%%%%%%%%
\section{Likelihood Analysis}\label{sec:3}
%%%%%%%%%%%%%%%%%%%%%%%%%%%%%%%%%

\begin{figure}[t]
\centering
\includegraphics[width=9cm]{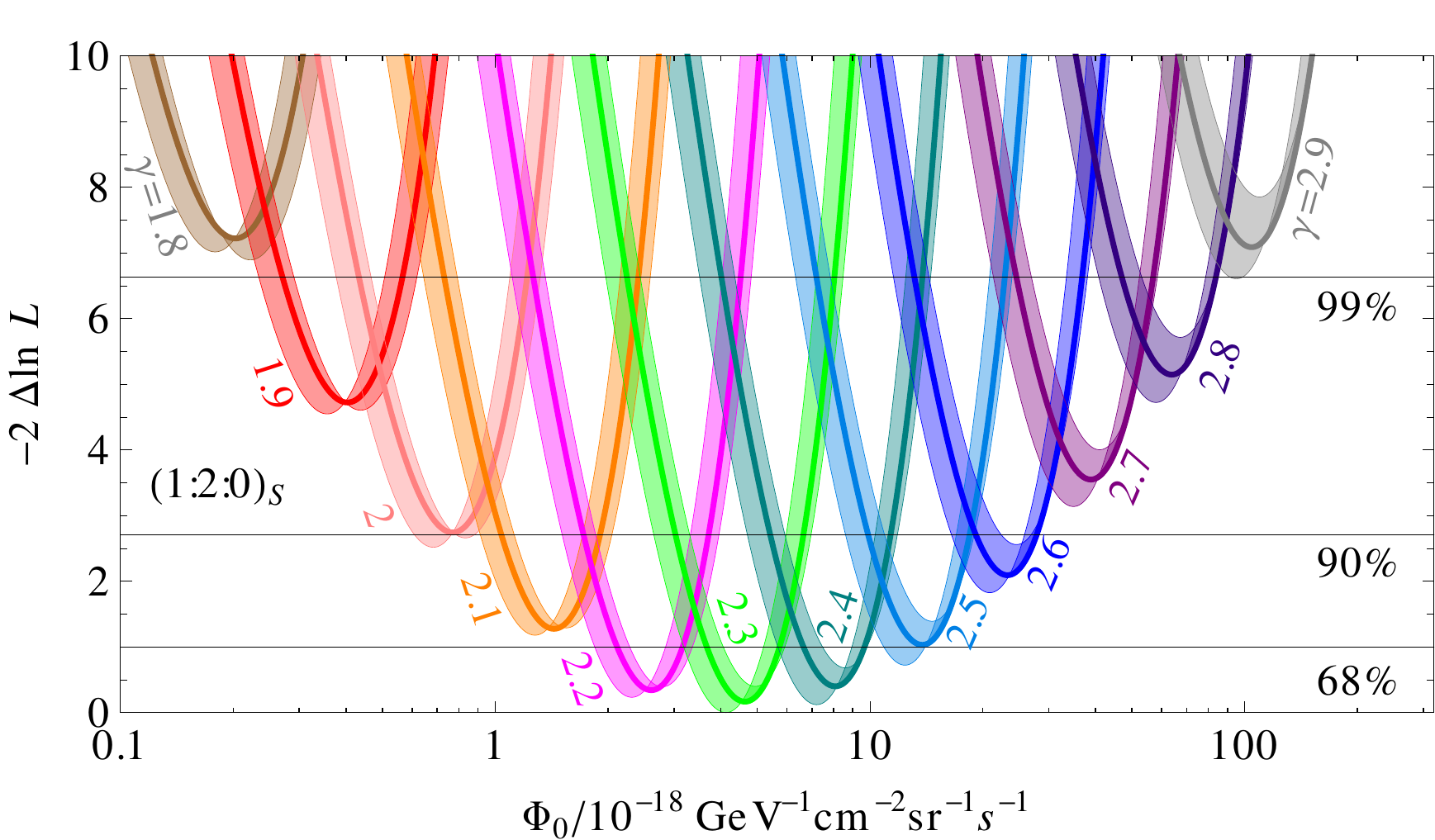}
\caption{The likelihood profile for the total astrophysical flux normalization $\Phi_0$ with different values of the spectral index $\gamma$ and with (1:2:0)$_{\rm S}$ flavor composition. The shaded regions show the 90\% CL PDF uncertainty. } 
\label{fig:like1}
\end{figure}

%{\em $\chi^2$-Analysis.---}
Using the parameter values given above, we use Eq.~(\ref{eq:N}) to compute the expected number of events from the SM CC and NC interactions, assuming a single-component astrophysical power-law spectrum for the incoming neutrino flux. Together with the expected atmospheric background, we obtain the predictions for the SM signal+background events with mean values $\lambda_i$, where $i=1,...,14$ denotes the number of the deposited energy bin between $15.8~{\rm TeV} < E_{\rm dep} < 10~{\rm PeV}$. For a given flavor composition, the observed count $n_i$ in each bin $i$ is compared to the SM prediction through a Poisson likelihood function 
\begin{align}
L \ = \ \prod_{{\rm bins}~i} \frac{e^{-\lambda_i}\lambda_i^{n_i}}{n_i!}
\label{like}
\end{align}
The best-fit values for the flux normalization $\Phi_0$ and the spectral index $\gamma$ will correspond to the maximum value of $L=L_{\rm max}$ in Eq.~\eqref{like}. The corresponding confidence level (CL) ranges can be obtained from a likelihood-ratio test, by constructing a test statistic
\begin{align}
-2\Delta \ln L \ = \ -2(\ln L- \ln L_{\rm max}) \; ,
\end{align} 
where the 68\%, 90\% and 99\% CL limits correspond to the values of $-2\ln L = 1$, 2.71 and 6.63 respectively~\cite{pdg} for our case with one degree of freedom. This is shown in Figure~\ref{fig:like1} for the (1:2:0)$_{\rm S}$ flavor composition. Here the individual $L_{\rm max}$ was computed by only varying the spectral index and flux normalization for a given flavor composition. We find that the best-fit spectral index is $\gamma\sim 2.3$ and the best-fit normalization is $\Phi_0\sim 5\times 10^{-18}~{\rm GeV}^{-1}{\rm cm}^{-2}{\rm sr}^{-1}{\rm s}^{-1}$. Note that the flux normalization values shown here are consistent with the observational upper bounds on the UHECR and diffuse neutrino fluxes~\cite{limit}, as well as with the sub-PeV gamma-ray flux limits $E_\gamma^2 \Phi_\gamma \lesssim 10^{-8}~{\rm GeV}{\rm cm}^{-2}{\rm sr}^{-1}{\rm s}^{-1}$~\cite{Murase:2014tsa}. This might indicate a relation between the sources of the UHE neutrinos and UHECRs~\cite{Anchordoqui:2013dnh}. 
%If this is true, then the non-detection of UHE neutrinos beyond 2 PeV suggests a suppression of UHECR spectrum beyond $\sim$ 100 PeV, which may be due to interactions of UHECRs en route to Earth or due to a natural acceleration endpoint~\cite{Murase:2014tsa}. 
However, our best-fit spectral index is slightly higher than that predicted by a typical Waxman-Bahcall flux, which is proportional to $E^{-2}$~\cite{wb}.\footnote{This is largely consistent with other recent analyses of the IceCube data; see e.g.~\cite{Mena:2015mka, Aartsen:2015ivb, ic4, Kalashev:2014vra}.}

\begin{figure}[t]
\centering
\includegraphics[width=8cm]{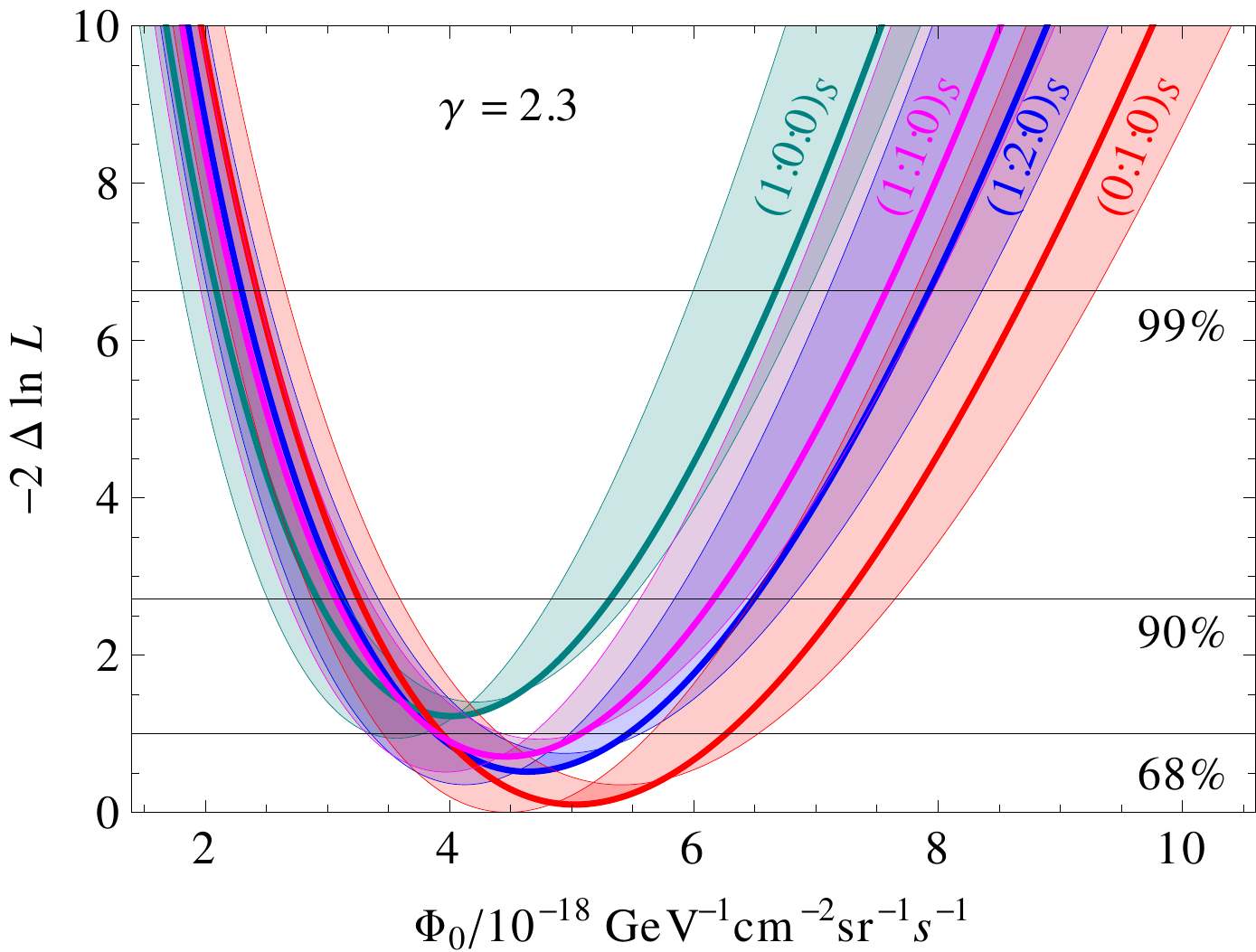}
\caption{The likelihood profile for the total astrophysical flux normalization $\Phi_0$ with different flavor compositions and with the spectral index $\gamma=2.3$. The shaded regions show the 90\% CL PDF uncertainty.}\label{global}
\end{figure}

Using the above procedure, we can similarly compute the likelihood profiles for other flavor compositions mentioned at the end of Section~\ref{sec:2}, which look similar to that shown in Figure~\ref{fig:like1}, and hence, are not shown here. We find that for all the cases, the best-fit spectral index is $\gamma=2.3$ and the best-fit flux normalization is of the same order as in the (1:2:0)$_{\rm S}$ case, though the exact value depends on the flavor composition. This is shown in Figure~\ref{global}. Here the overall maximum likelihood was computed by comparing all the corresponding local maxima $L_{\rm max}$ for each flavor composition. As we can see from this plot, all the physical flavor compositions are currently consistent with the IceCube data within 90\% CL, although the (0:1:0)$_{\rm S}$ flux has the maximum likelihood to explain the current data. This is consistent with the recent IceCube analysis on flavor composition~\cite{Aartsen:2015ivb}. 

\begin{figure*}[t!]
\centering
\includegraphics[width=8cm]{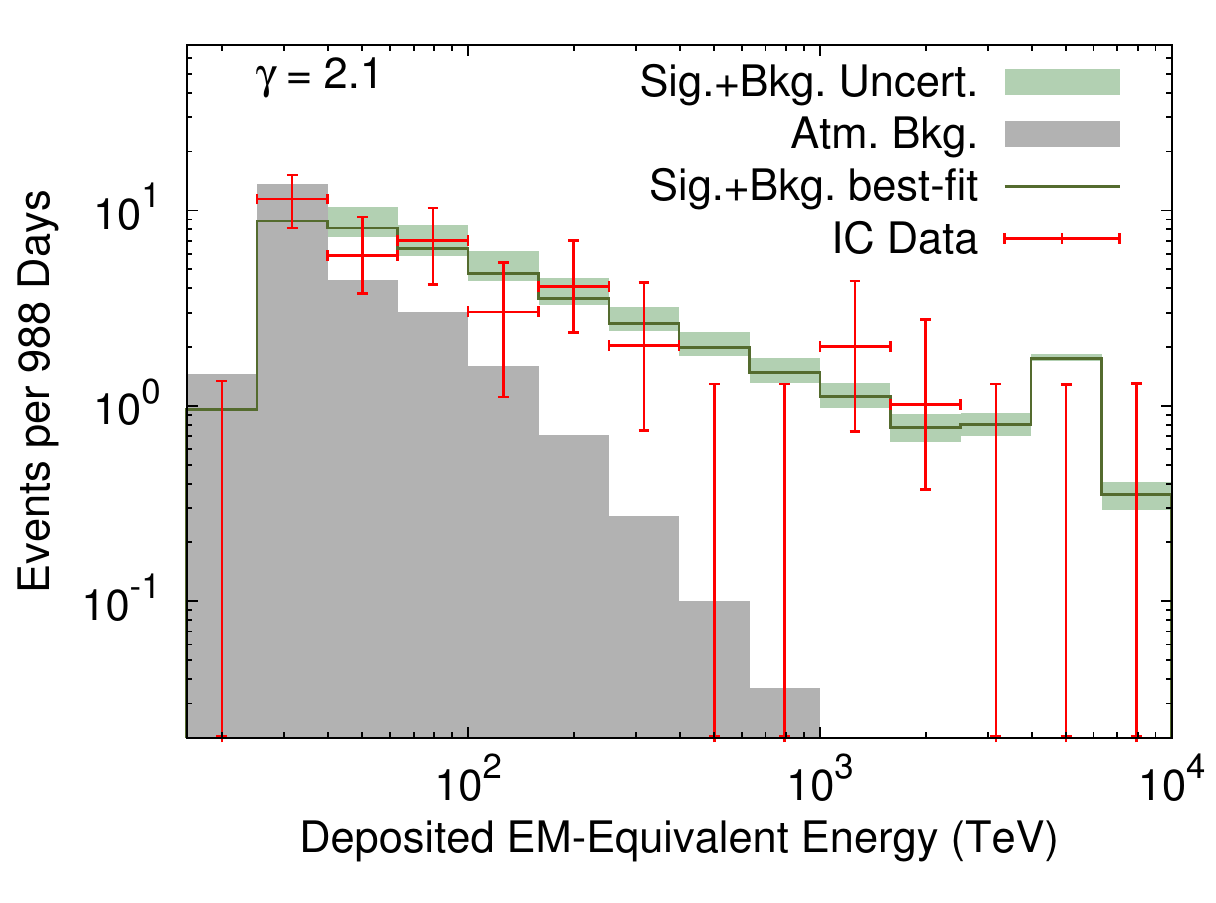} \hspace{0.5cm}
\includegraphics[width=8cm]{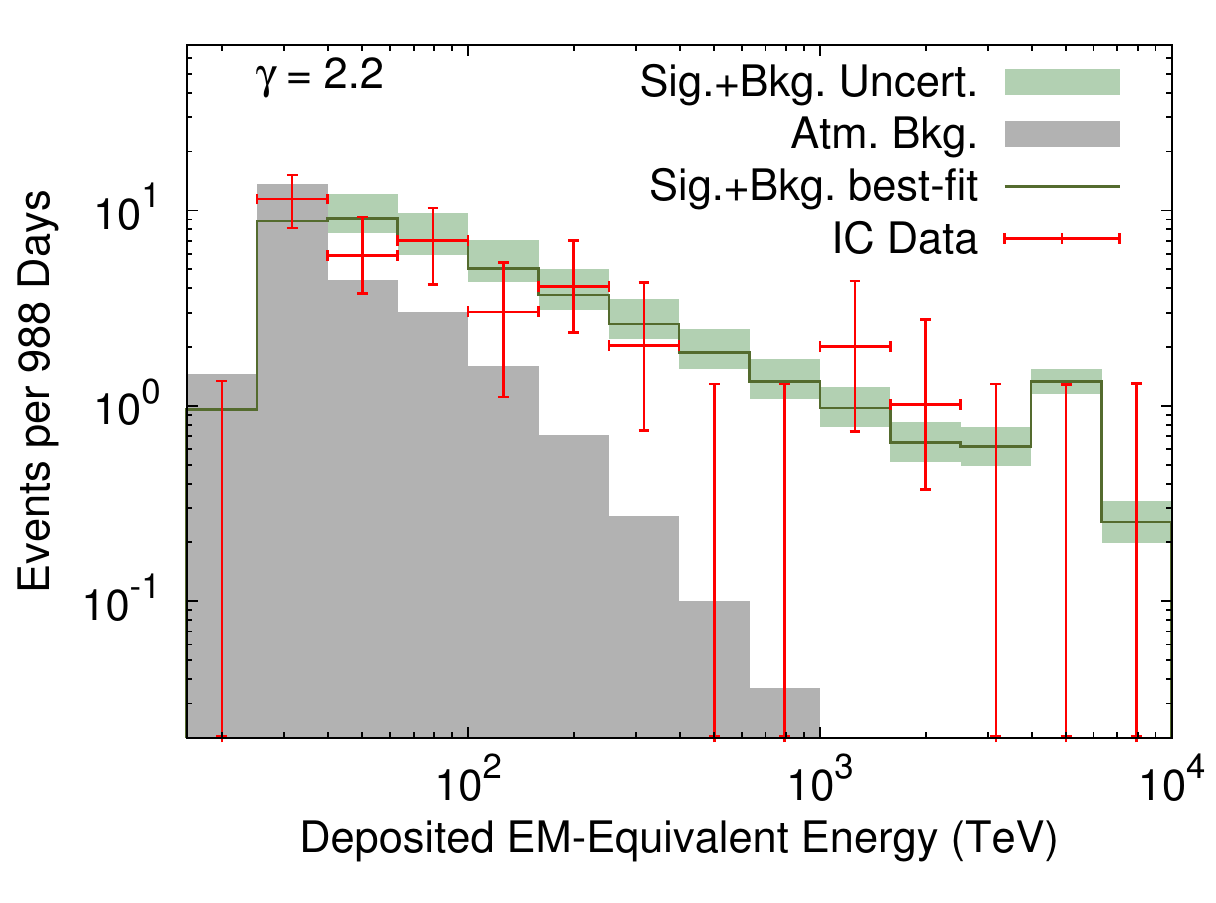} \\
\includegraphics[width=8cm]{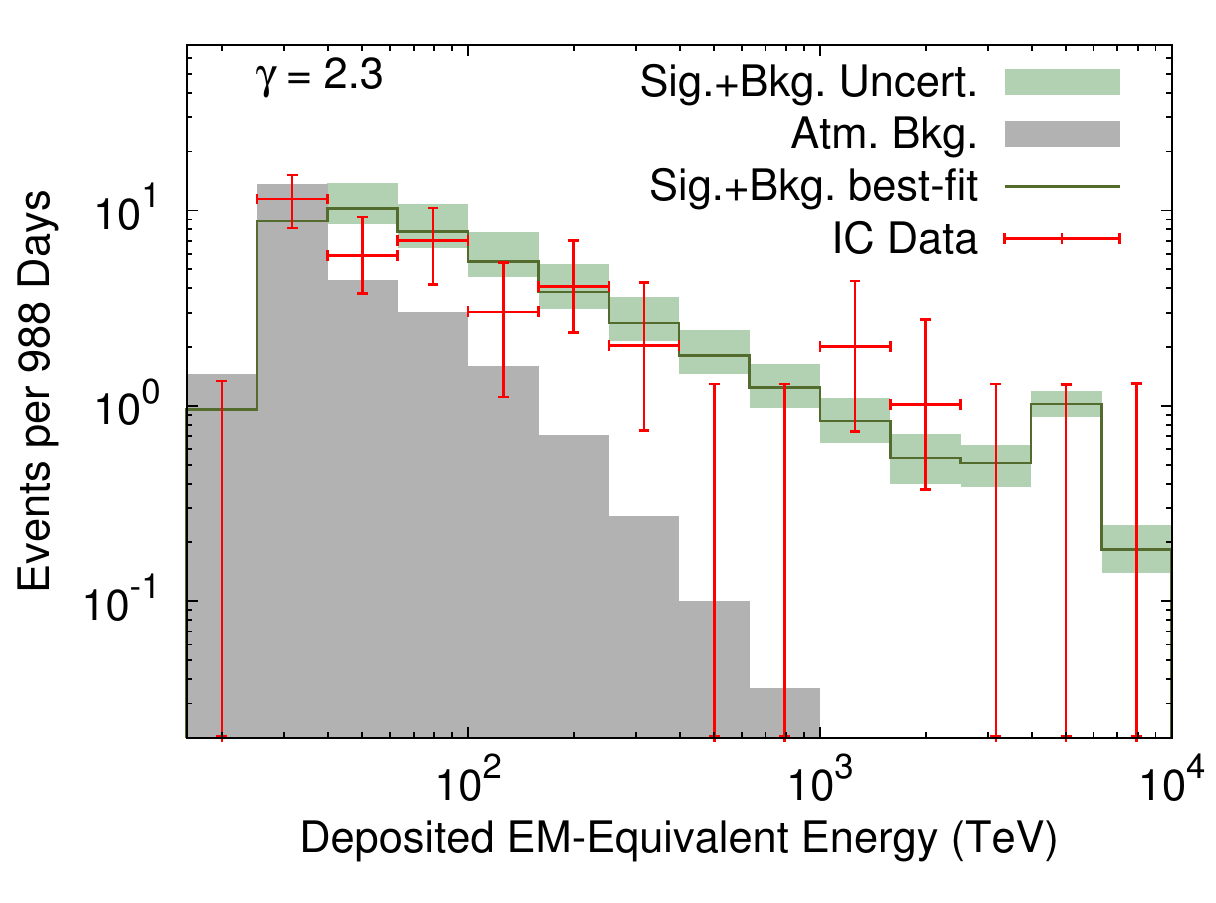}\hspace{0.5cm}
\includegraphics[width=8cm]{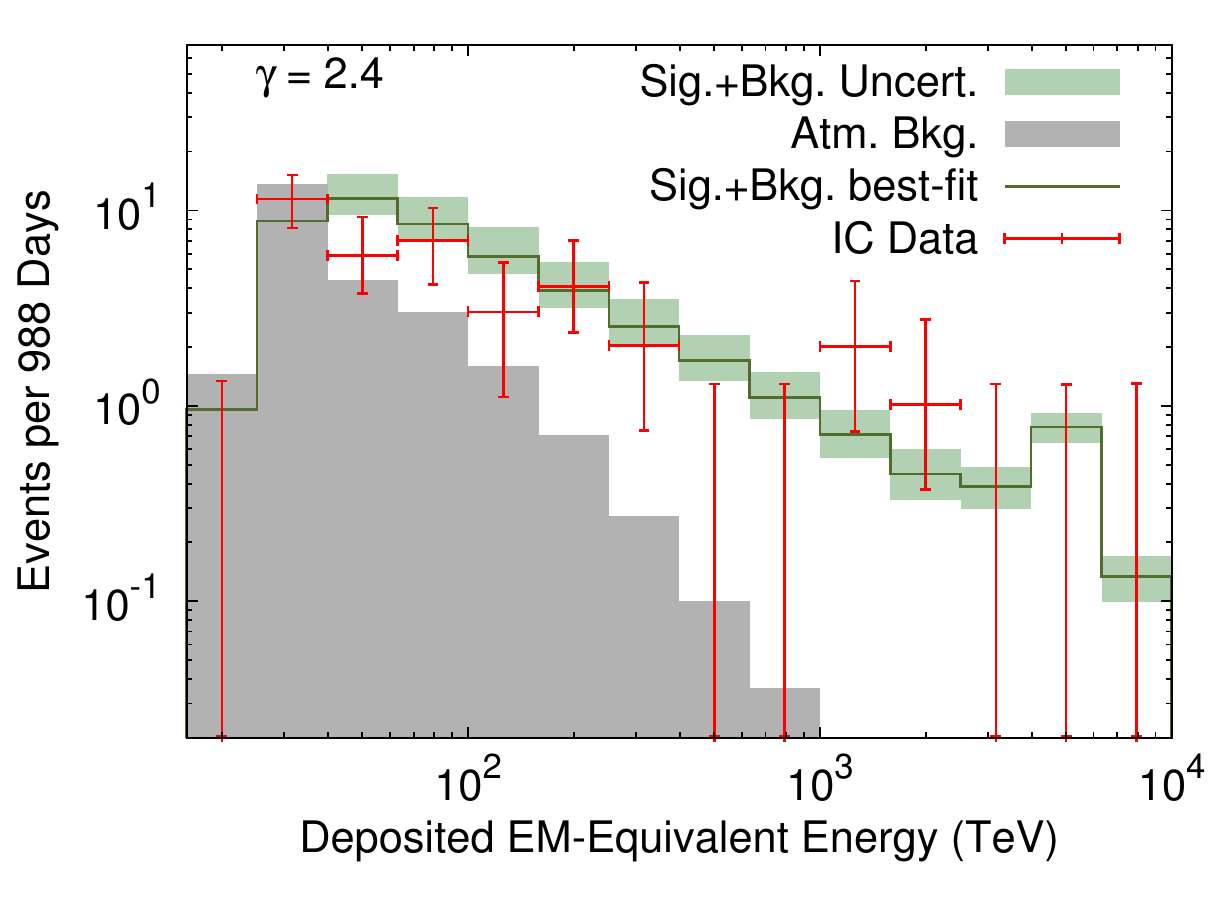}\\
\includegraphics[width=8cm]{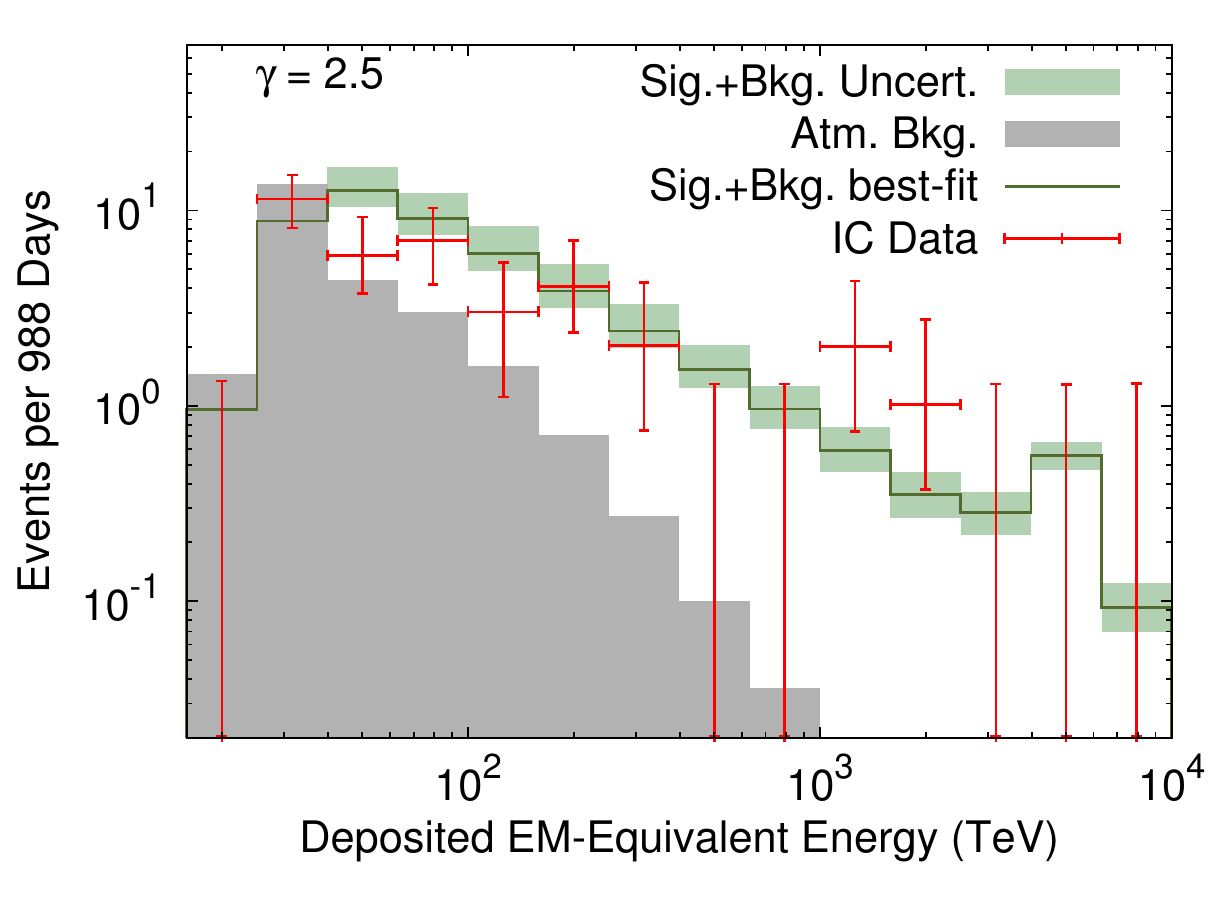}\hspace{0.5cm}
\includegraphics[width=8cm]{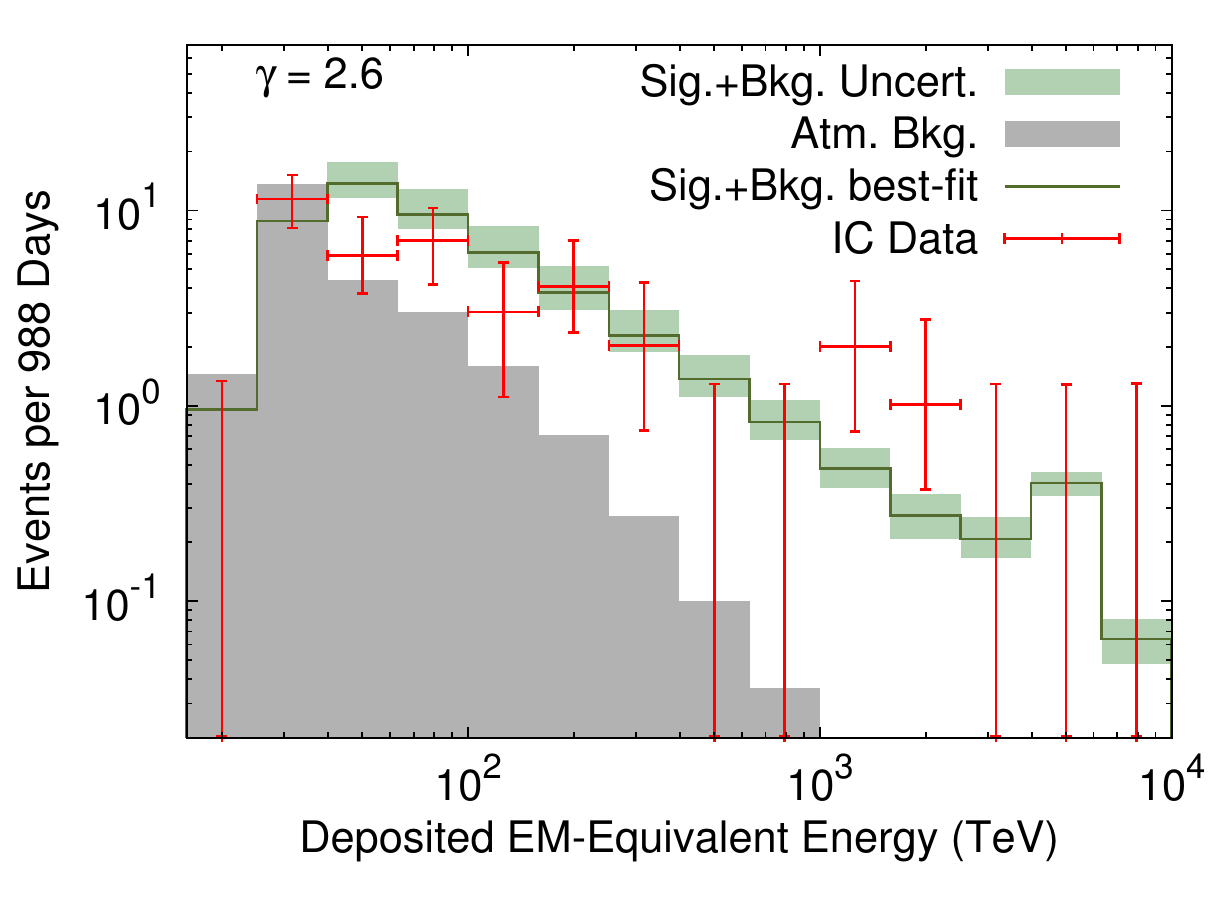}\\
\includegraphics[width=8cm]{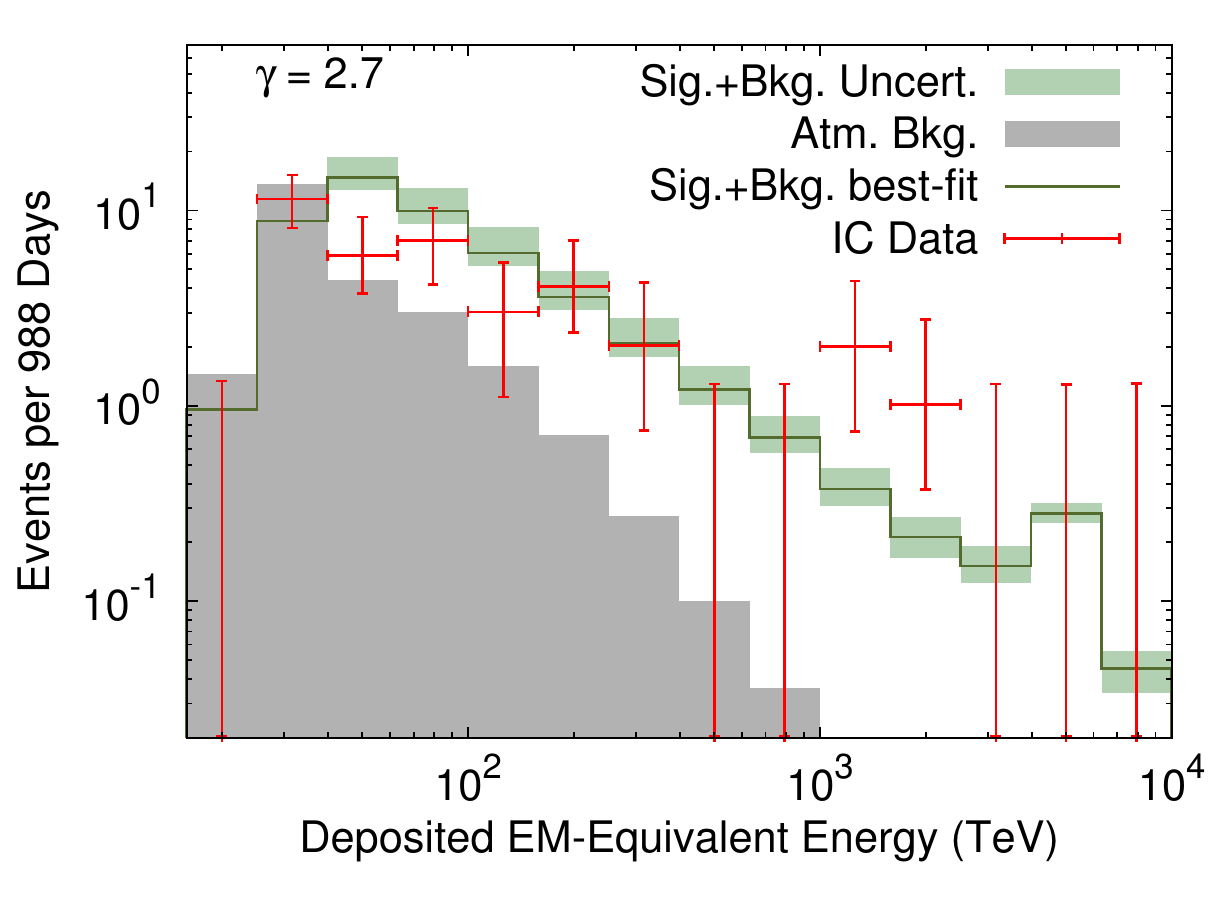}\hspace{0.5cm}
\includegraphics[width=8cm]{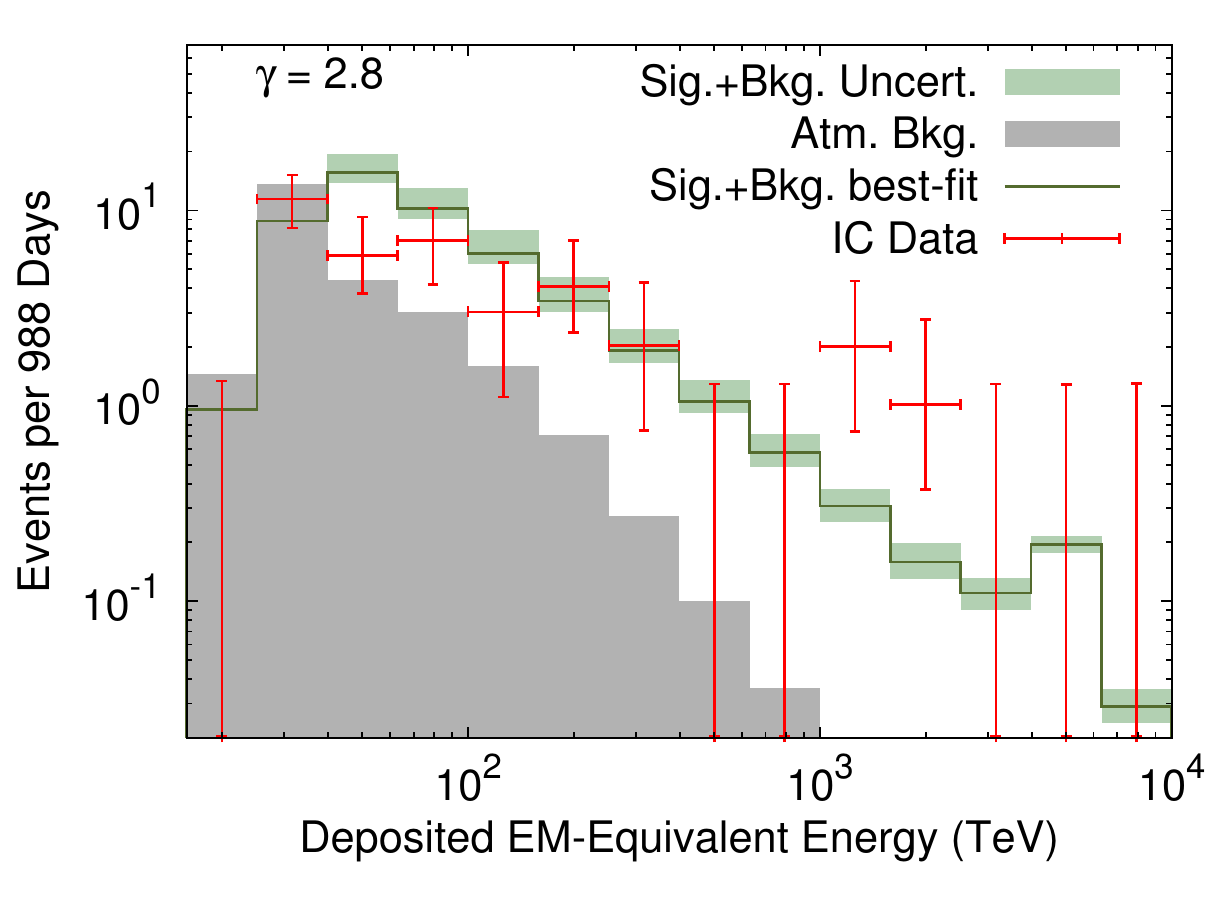}
\caption{The SM signal+background events for an $E^{-\gamma}$ flux with (1:1:1)$_{\rm E}$ flavor composition, along with their $3\sigma$ uncertainties (green shaded), for the IceCube deposited energy bins between 16 TeV - 10 PeV. The IceCube data points (with error bars) and the 
atmospheric background (black shaded) 
were taken from~\cite{ic3}.} %These distributions along with the corresponding ones for a (2:1:1)$_{\rm E}$ composition (not shown here) are used to derive the allowed regions shown in Figure~\ref{fig1}. }
\label{fig2}
\end{figure*} 

\begin{table*}[t]
\begin{center}
\begin{tabular}{c|c|c|c|c|c|c|c}\hline\hline
 & Background & (1:2:0)$_{\rm S}$ & (1:1:0)$_{\rm S}$ & (0:1:0)$_{\rm S}$ & (1:0:0)$_{\rm S}$ & Two-comp & IceCube \\ 
& & [or (1:1:1)$_{\rm E}$] & [or (14:11:11)$_{\rm E}$]  & [or (4:7:7)$_{\rm E}$] & [or (5:2:2)$_{\rm E}$] & [(1:0:0)$_{\rm S}+$(1:2:0)$_{\rm S}$] & \\ \hline\hline
Total & $2.8+<5.3$ & 14.5 & 14.4 & 14.8 &  14.2 &  13.7 & 20 \\
Up & $1.5+<3.7$ & 5.5 & 5.4 & 5.6 & 5.2 & 5.0  & 5 \\
Down & $1.2+<1.6$ & 9.0 & 9.0 & 9.1 & 9.0 & 8.7 & 15\\
Track & $\sim 2.1+<1.0$ & 4.4 & 3.9 & 5.6  & 2.6 & 2.8 & 4\\
Shower & $\sim 0.7+<4.2$ & 10.1 & 10.5 & 9.2 & 11.6  & 10.9 & 16\\ \hline
%$p$-value & & 0.95 & 0.95 & 0.75 & \\ \hline 
\hline
\end{tabular}
\end{center}
\caption{SM predictions for the number of events between 60 TeV $< E_{\rm dep}<$ 3 PeV in 988 days  for the best-fit single-component solutions as well as for a two-component solution. The atmospheric background due to CR muons and muon neutrinos from $\pi/K$ and charmed meson decays, and the IceCube observed events are taken from~\cite{ic3}. } \label{tab1}
\end{table*}

In Figure~\ref{fig2}, we show the deposited energy spectra for a single component flux (\ref{phi}) with (1:1:1)$_{\rm E}$ flavor composition and an unbroken $E^{-\gamma}$ spectrum for some typical  values of the spectral index $\gamma$ . The expected background of $6.6^{+5.9}_{-1.6}$ atmospheric neutrinos and $8.4\pm 4.2$ CR muons is shown by the black shaded region, which includes the systematic and statistical uncertainties as well as the 90\% CL charm limit~\cite{ic3}. %\footnote{Recently, it has been suggested~\cite{Gaisser:2014pda} that for energies above 100 TeV, there might be a small ($\sim 10\%$) increase in the atmospheric $\nu_e$ flux from $K_S\to \pi e \nu_e$.} Note that 
%Since the IceCube data in the first two bins are compatible with the atmospheric background, we have assumed the astrophysical flux to dominate after 60 TeV deposited energy.  
Our SM signal+background prediction is shown by the green solid line and the associated green shaded region includes the PDF uncertainty in the cross section (cf.~Figure~\ref{fig2}). The mild enhancement of events around 6 PeV is due to the Glashow resonance caused by $\bar{\nu}_e e^-$ interactions~\cite{glashow}.

%From Figure~\ref{fig2}, the SM expected number of events with $(1:1:1)_E$ flavor composition seems to be consistent with the IceCube data within their current uncertainties. 
%However, 
A closer look at the signal events for (1:2:0)$_{\rm S}$ reveals a potential `muon deficit' problem, as illustrated in  Table~\ref{tab1}. Here we have shown the expected number of signal events for the best-fit scenario with $\gamma=2.3$ and (1:2:0)$_{\rm S}$ composition in various categories. Including the contribution from the expected atmospheric background~\cite{ic3} due to CR muons and atmospheric muon neutrinos from $\pi/K$ and charmed meson decays, we compare our best-fit predictions with the observed data for integrated number of events between 60 TeV $< E_{\rm dep} < $ 3 PeV. It is clear that the (1:2:0)$_{\rm S}$ flavor composition predicts higher number of muon tracks than that observed in this energy range. 
%Note that our predicted track-to-shower ratio of $\sim 20\%$ is consistent with earlier results
%~\cite{candia}. 
This deficit becomes more severe if we consider the fact that out of the 4 candidate events above 60 TeV, one event has an apparent first interaction near the detector boundary, which is consistent with the expected muon background~\cite{ic3}. %Thus, {\it only} 3 events might be the candidate signal track events above 60 TeV. 
%Including the through-going muons with their interaction vertex outside the detector volume in the analysis might alleviate this problem. However, in the absence of sufficient experimental information on this issue, we look for an alternative possible solution. 

For comparison, we also show in Table~\ref{tab1} the corresponding predictions for number of events with other physical flavor compositions. It is clear that the muon deficit problem becomes worse for (1:1:0)$_{\rm S}$ and (0:1:0)$_{\rm S}$, whereas (1:0:0)$_{\rm S}$ significantly improves the situation, as expected.  
%In the following section, we provide a physical motivation for the (1:0:0)$_{\rm S}$ flux. 
%Here we wish to comment that the existence of the gap is not statistically significant in the current data. Nevertheless, we must emphasize that a clear evidence of the gap in future data might 

As suggested in~\cite{Chen:2013dza}, one possible way to address the muon deficit problem is by invoking some exotic lepton flavor violating interactions, which could also be linked with the longstanding muon $(g-2)$ anomaly~\cite{pdg}. 
%However, this requires an asymmetric flavor-violating couplings of the new gauge boson $Z'$ to quarks and/or leptons~\cite{Murakami:2001cs}, which is difficult to embed in a realistic model. 
It was shown~\cite{Araki:2014ona} that a light leptophilic $Z'$ explaining the muon $(g-2)$ anomaly could also explain the apparent energy gap between 400 TeV and 1 PeV in the IceCube spectrum as due to resonant scattering of UHE neutrinos with the cosmic relic neutrino background. 
%A UV-complete model for such light $Z'$ was recently constructed in~\cite{Kamada:2015era}. 
In the following section, we propose an alternative explanation of the gap within the SM framework. 
%Here we wish to comment that the existence of the gap is not statistically significant in the current data, as can be seen from Figure~\ref{fig2}. However, a clear evidence of the gap in future data might suggest the existence of new physics or a secret neutrino interaction with a light ($\sim$ MeV-scale) mediator~\cite{blum, Araki:2014ona, gap}. 
%%%%%%%%%%%%%%%%%%%%%%%%%%%

\section{A Two-Component Solution} \label{sec:5}
%%%%%%%%%%%%%%%%%%%%%%%%%%%%%%%%%%%%%%%

%{\em A Two-Component Flux.---}
As stated earlier, once we identify the existence of more than one flavor compositions, it seems rather natural to consider a multi-component flux for the astrophysical neutrinos. Here we should clarify that multiple flavor components do not necessarily require multiple sources, since it is possible to have significant parameter spread even in a single source class. For example, the energy at which muons of astrophysical origin start to lose energy before decaying strongly depends on the source parameters such as the magnetic field strength. For a given source class, such parameters could have a broad distribution, thus leading to a diffuse neutrino flux with more than one flavor composition.

For illustration, we entertain the simplest multi-component possibility, i.e. a two-component flux: 
\begin{eqnarray}
\Phi(E_\nu) \ = \  \Phi_1 \left(\frac{E_\nu}{E_0}\right)^{-\gamma_1} e^{-E_\nu/E_1}+ \Phi_2 \left(\frac{E_\nu}{E_0}\right)^{-\gamma_2} ,
\label{phi2}
\end{eqnarray}
with five hitherto unknown parameters, i.e. $\Phi_{1,2}, \gamma_{1,2}$ and a cut-off scale $E_1$. Note that in the single-component case, there is no need for a cut-off~\cite{Chen:2013dza, Kalashev:2014vra}, as long as $\gamma\gtrsim 2$.\footnote{For $\gamma\lesssim 2$, a cut-off is required to avoid the Fermi-LAT constraint on diffuse $\gamma$-ray flux~\cite{Murase:2014tsa}.}  However, if the apparent gap in the deposited energy spectrum between 400 TeV and 1 PeV persists, the single-component solution will be disfavored, as illustrated in Figure~\ref{fig2}. A two-component flux (\ref{phi2}) can easily explain such a gap, if $E_1\lesssim 1$ PeV for the first component and the second component becomes dominant beyond $E_1$. The cut-off energy $E_1$  may arise due to interactions of UHECRs en route to Earth or due to a natural acceleration endpoint~\cite{Murase:2014tsa}. Similarly, a low-energy cut-off for the second component could be due to its production mechanism, e.g. from a $p\gamma$ collision which has a sharp $\Delta$-resonance.\footnote{In general, it is possible to back up such a two-component flux \eqref{phi2} by solid astrophysical scenarios; a detailed discussion of this will be postponed to a future work. Another possibility for the two-component flux is that one of the components (preferably the lower one) could have astrophysical origin, while the other one has some exotic origin, such as a decaying DM scenario~\cite{Feldstein:2013kka}.} %Similarly, the second component at energies above $E_1$ could have an initially rising energy spectrum due to a spin-flip mechanism for the photons, analogous to that in the $B$-meson decay. 

Similarly, the potential muon deficit problem can be adressed by taking the first component of the two-component flux to be (1:0:0)$_{\rm S}$. We should clarify that although the (1:0:0)$_{\rm S}$ is not favored by the current data over the other possible flavor compositions as a {\em single}-component flux, mainly because of the lack of enough events in the Glashow resonance bin, it is still perfectly acceptable as the low-energy part of a two-component flux. Moreover, it is easier to satisfy the Fermi-LAT bound on diffuse gamma-ray flux~\cite{Ackermann:2014usa} for a (1:0:0)$_{\rm S}$ source in the low-energy regime~\cite{Anchordoqui:2014pca}.

%Due to the lack of enough statistics at the moment, it is difficult to distinguish the two-com scenario from a single-component flux or to constrain all four parameters in (\ref{phi2}). However, the two-component flux (\ref{phi2}) could explain some features of the current data better than the single-component flux. To illustrate this point further, we consider the situation where {\em both} $(1:1:1)_E$ and $(2:1:1)_E$ fluxes are present simultaneously, but in different energy regimes separated by the apparent gap in the deposited energy spectrum between 300 TeV and 1 PeV. As we show below, this provides an alternative explanation of the gap within the SM interaction framework. 

\begin{figure}[t]
\centering
\includegraphics[width=8cm]{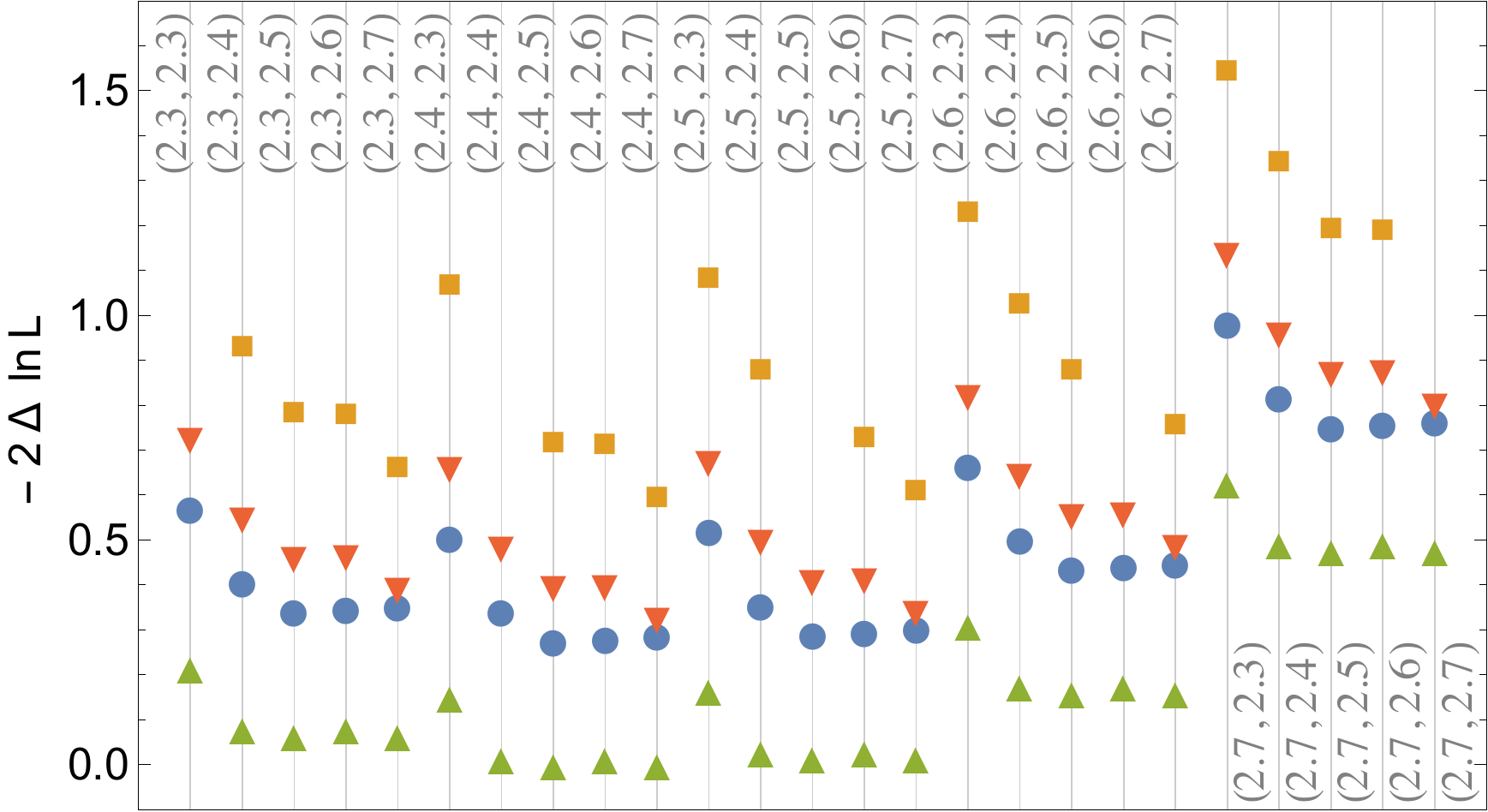}
\caption{A likelihood analysis for the two component flux for different flavor compositions, the first component being (1:0:0)$_{\rm S}$ and the second being (1:2:0)$_{\rm S}$ (circles), (1:0:0)$_{\rm S}$ (squares), (0:1:0)$_{\rm S}$ (triangle up) and (1:1:0)$_{\rm S}$ (triangle down). The $x$-axis corresponds to the variation of the spectral indices $(\gamma_1,\gamma_2)$, as indicated in the figure.}
\label{fig4}
\end{figure}

Assuming the first component to be (1:0:0)$_{\rm S}$ for reasons stated above, we perform a likelihood analysis for the two-component case, similar to that presented in Section~\ref{sec:3} for the single-component case, with different choices of flavor compositions for the second component and different spectral indices $(\gamma_1,\gamma_2)$ in Eq.~\eqref{phi2}.  We keep the cut-off scale fixed at $E_1=1$ PeV. Our results are shown in Figure~\ref{fig4}, where we show the relative likelihood for each case. It is clear that the best-fit is obtained for $(\gamma_1,\gamma_2)=(2.4-2.5,2.4-2.7)$ and for the second component flavor composition  (0:1:0)$_{\rm S}$. However, the other cases shown in Figure~\ref{fig4} are all currently allowed  within $90\%$ CL, if we consider the UHE neutrino events only above 60 TeV. This is consistent with the fact that, for the single-component case, (0:1:0)$_{\rm S}$ gives the best-fit solution, while all the other cases are still within the 90\% CL interval (cf. Figure~\ref{global}) for $\gamma=2.3-2.7$. 

\begin{figure}[t]
\centering
\includegraphics[width=8cm]{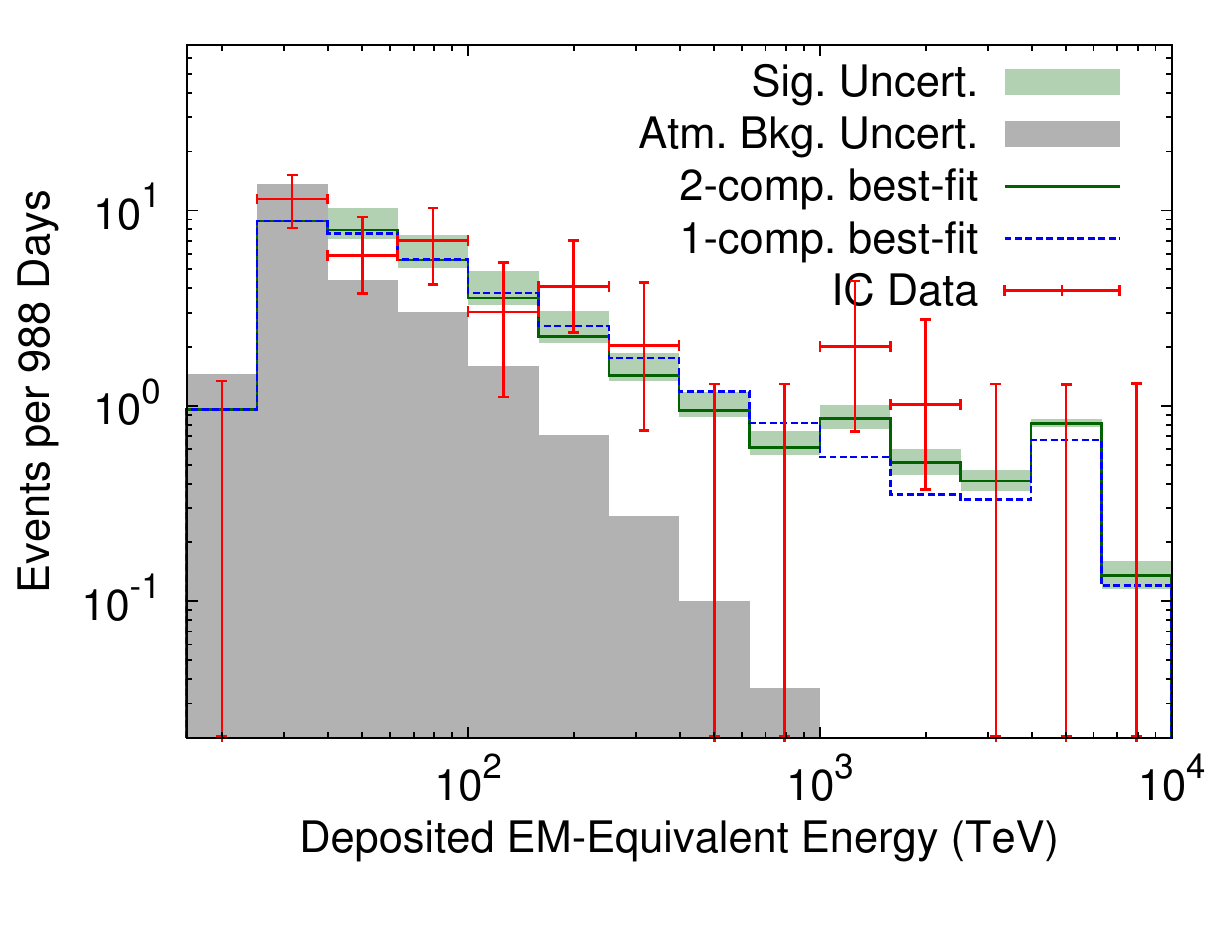}
\caption{The SM signal+background events for a two-component flux (solid line with $90\%$ CL PDF uncertainty band) with (1:0:0)$_{\rm S}$ as the low-energy component and (1:2:0)$_{\rm S}$ as the high-energy component. %having (2:1:1)$_{\rm E}$ with $\gamma_1=2.6$ and (1:1:1)$_{\rm E}$ with $\gamma_2=2.2$. 
We also show the best-fit single-component (1:2:0)$_{\rm S}$ flux for comparison. The IceCube data points (with error bars) and the 
atmospheric background (black shaded) 
were taken from~\cite{ic3}. }
\label{fig3}
\end{figure} 

To illustrate the effect of the two-component flux on the event distribution, as compared to the single component flux shown in Figure~\ref{fig2}, we change the low-energy part of the flavor composition to (1:0:0)$_{\rm S}$, but keeping the high-energy part same as in Figure~\ref{fig2}, i.e. (1:2:0)$_{\rm S}$. We keep the cut-off scale fixed at $E_1=1$ PeV and use the best-fit solution for this flavor composition from Figure~\ref{fig4}, i.e. $(\gamma_1, \gamma_2)=(2.4,2.5)$, with the corresponding best-fit normalizations $(\Phi_1,\Phi_2)=(6.4,32.2)\times 10^{-18}~{\rm GeV}^{-1}{\rm cm}^{-2}{\rm sr}^{-1}{\rm s}^{-1}$.  The resulting number of events for this scenario are given in the penultimate column of Table~\ref{tab1}, which clearly shows that the muon deficit can be easily addressed in this scenario. The energy spectrum for this two-component solution is shown in Figure~\ref{fig3} along with the 90\% CL PDF uncertainty (green shaded band). The best-fit (1:2:0)$_{\rm S}$ single-component solution is also shown for comparison. 
%Note that a harder (softer) spectral index for the lower (higher) energy component gives a better fit to the data, as explicitly demonstrated in Figure~\ref{fig2}. 
We see that the two-component flux better explains the energy gap just below 1 PeV. With more data in future, the two-component flux can in principle be distinguished from a single-component flux. In particular, any further information on the number of events in the Glashow resonance bin will be crucial to pin down the flavor composition in the higher-energy bins. In the two-component example considered above, the (1:2:0)$_{\rm S}$ flux in the higher energy range predicts less number of electron neutrino events in the Glashow resonance bin, and therefore, is preferred by the current data over a (1:0:0)$_{\rm S}$ flux. In addition, a more accurate measurement of the energy spectrum can distinguish our two-component hypothesis from other possible explanations of the gap, e.g. due to secret neutrino  interactions~\cite{nuSI, Araki:2014ona} or due to the line-of-sight interactions of CRs emitted by blazars with background photons~\cite{Essey:2009ju}.

%In fact, since the $(2:1:1)_E$ flavor composition gives less number of muon tracks, and the number of contained muon tracks is expected to be much smaller at energies above $\sim$ 300 TeV, one could envisage a two-component solution with the $(2:1:1)$ component contributing to the lower energy bins, and the $(1:1:1)$ component to the higher energy bins. This kind of solution can be tested in future with more statistics, as the number and energy distribution of expected muon tracks would be different in this case from the other scenarios discussed above. 

%%%%%%%%%%%%%%%%%%%%%%%%%%%%%%%%%%%%%
\section{Conclusion} \label{sec:6}
%%%%%%%%%%%%%%%%%%%%%%%%%%%%%%%%%%
%{\em Summary.---}
Understanding all aspects of the IceCube UHE neutrino events is extremely important for both  astrophysics and particle physics.  
%We have performed a spectral and flavor analysis of the 3-year IceCube dataset to obtain the 
%allowed range of the assumed astrophysical neutrino flux parameters. We stress that 
%The current data seems to be largely consistent with the SM expectations within the known uncertainties, without requiring any exotic explanation. With more statistics, this could provide a unique test of the SM interactions up to the highest energies ever observed in Nature, and in the absence of any significant deviations from the observed data, could be used to constrain various beyond SM scenarios. Nevertheless, 
We critically examine the single-component hypothesis with the standard (1:2:0)$_{\rm S}$ astrophysical neutrino flux from pion/kaon decays. Identifying the existence of other well-motivated physical flavor compositions,  we argue that it is natural to have a multi-component flux rather than a single-component one. We further show that a simple two-component flux could explain all 
the key features of the data {\em within} the SM framework, which are otherwise difficult to understand with a single-component flux. In particular, if the apparent energy gap between 400 TeV and 1 PeV deposited energy bins becomes statistically significant, our two-component hypothesis might provide a simple explanation for this observation, with important consequences for the identification of the underlying astrophysical sources. 
%The two-component flux proposed here can in principle be distinguished from a single-component flux with more data in future. 
Given that so much is unknown about the dynamics of the UHE neutrino sources and that a precise knowledge of their flavor composition is crucial for a reliable understanding of the underlying astrophysical processes as well as for the particle physics interpretation, it is desirable to determine the flavor composition from experiment, as more data is collected at current and future large-volume neutrino detectors.  Finally, we hope that the future experimental analyses will seriously consider the possibility of such a two-component flux.   
%Therefore, it is extremely important to extract the flavor composition of the astrophysical neutrino flux directly from the data, rather than just assuming a particular flavor ratio, in order to understand the origin of the UHE neutrinos. 
%Possible implications of our analysis for various new physics scenarios will be explored in a follow-up communication. 

\section*{Acknowledgments}
%%%%%%%%%%%%%%%%%%%%%%%%%%%%%%%
%{\em Acknowledgments.---}
We thank Alex Kusenko, Kohta Murase and Sergio Palomares-Ruiz for helpful comments and discussions. The work of P.S.B.D. is supported by
 the Lancaster-Manchester-Sheffield Consortium for Fundamental
 Physics under STFC grant ST/L000520/1.  The work of C-Y.C. and A.S. is supported in part 
by the US Department of Energy under Grant DE-AC02-98CH10886. P.S.B.D. acknowledges the local hospitality at BNL where part of this work was done. 
 
%%%%%%%%%%%%%%%%%%%%%%%%%%%%%%%%%%%%%%%%%%%%%%%%%%%%%%%%%%%%
\section*{Note Added}
%{\em Note Added.---} 
After this work was finalized, we became aware of the preliminary 4-year IceCube dataset~\cite{botner}, in which 
%Some of the key features of the data, as described in this work as motivations for a two-component flux, still persist in the new dataset. 
%the observed track-to-shower ratio at high energies no longer suggests a muon deficit. However, 
the apparent energy gap between 400 TeV - 1 PeV still seems to be present, as the only new event recorded in this range is a muon track, which is most likely to have originated from a neutrino with much higher incoming energy. %Thus, our two-component hypothesis still provides a simple explanation of the gap, though it may no longer be necessary to invoke a (1:0:0)$_{\rm S}$ component. In any case, 
We look forward to the formal publication of the new dataset with more information on the events as well as on the atmospheric background, to be able to update our spectral and flavor analysis accordingly in a future work.

\end{document}